\documentclass[%
aip,
reprint,
 amsmath,amssymb,
superscriptaddress,
]{revtex4-1}

\usepackage{color} 
\usepackage{amsmath}
\usepackage{natbib}
\usepackage[export]{adjustbox}
\usepackage{graphicx}
\usepackage{dcolumn}
\usepackage{bm}
\usepackage{subfigure}

\usepackage[utf8]{inputenc}
\usepackage[T1]{fontenc}
\usepackage{mathptmx}

\begin{document}


\title{Distinguishing resonance symmetries with energy-resolved photoion angular distributions from ion-pair formation in O$_2$ following two-photon absorption of a 9.3~eV femtosecond pulse}

\author{Kirk A. Larsen}
\email{klarsen@lbl.gov}
\affiliation{%
 Graduate Group in Applied Science and Technology, University of California, Berkeley, CA 94720, USA}
\affiliation{%
 Chemical Sciences Division, Lawrence Berkeley National Laboratory, Berkeley, CA 94720, USA}%
 
\author{Robert R. Lucchese}
\affiliation{%
 Chemical Sciences Division, Lawrence Berkeley National Laboratory, Berkeley, CA 94720, USA}%
 
\author{Daniel S. Slaughter}
\affiliation{%
 Chemical Sciences Division, Lawrence Berkeley National Laboratory, Berkeley, CA 94720, USA}%
 
\author{Thorsten Weber}
\email{tweber@lbl.gov}
\affiliation{%
 Chemical Sciences Division, Lawrence Berkeley National Laboratory, Berkeley, CA 94720, USA}%

\date{\today}

\begin{abstract}
We present a combined experimental and theoretical study on the photodissociation dynamics of ion-pair formation in O$_2$ following resonant two-photon absorption of a 9.3 eV femtosecond pulse, where the resulting O$^+$ ions are detected using 3-D momentum imaging. Ion-pair formation states of $^3\Sigma^-_g$ and $^3\Pi_g$ symmetry are accessed through predissociation of optically dark continuum Rydberg states converging to the B $^2\Sigma^-_g$ ionic state, which are resonantly populated via a mixture of both parallel-parallel and parallel-perpendicular two-photon transitions. This mixture is evident in the angular distribution of the dissociation relative to the light polarization, and varies with the kinetic energy release (KER) of the fragmenting ion-pair. The KER-dependent photoion angular distribution reveals the underlying two-photon absorption dynamics involved in the ion-pair production mechanism and indicates the existence of two nearly degenerate continuum resonances possessing different symmetries, which can both decay by coupling to ion-pair states of the same total symmetry through internal conversion.
\end{abstract}

\pacs{Valid PACS appear here}
\maketitle


\section{\label{sec:level1}Introduction}

Discrete electronic states lying above the first ionization threshold are energetically degenerate with states that are members of a continuous spectrum. Such discrete continuum states were postulated in the early days of quantum mechanics by figures such as von Neumann and Wigner~\cite{Wigner}, and were later studied by Fano~\cite{Fano}. The configuration interaction between the discrete and continuous spectra causes such continuum-embedded discrete states to become quasi-bound and thus possess a finite lifetime. The coupling between the resonances and continuum creates a substantial challenge for theoretical descriptions of electronic structure, and on detailed experimental measurements to test those methods.

Many resonances with appreciable scattering amplitude may appear near-by one another energetically, so that the probability of interaction between the photoelectron and the core is highly dependent on the excitation energy. Neighboring continuum resonances may be interacting or non-interacting, depending on their relative positions and widths. If two adjacent resonances are broad, as in the case of shape resonances, they may overlap with each other. When such resonances possess different symmetries, yet combined with the core produce final states of the same total symmetry, the resonances can interfere and result in a rapid change in the electron emission pattern \cite{Larsen1}. It is therefore of fundamental interest to identify and characterize neighboring continuum resonances, to better understand their properties, and to distinguish dynamics from quantum inference effects.

Electron-ion resonances in molecules often possess different decay mechanisms that compete, some involving ionization (e.g. autoionization) and others where all electrons remain bound (e.g. predissociation). These processes involve many-electron dynamics and are strongly mediated by electron-electron correlation. In certain instances, discrete continuum states may be predissociated by states that fragment to a cation-anion pair, rather than a neutral pair. When such resonances lie below the first dissociative ionization threshold, they are uniquely situated to be studied through the fragment ions emerging from the decay to the aforementioned ion-pair states. This offers advantages over photoelectron detection, because photoelectrons originate from various channels, such that photoelectrons emerging from autoionization compete with a stronger signal of photoelectrons emerging from direct ionization, leading to overlapping or unresolved photoelectron spectral features. In contrast, the ionic fragments emerge from the ion-pair formation channels, as dissociative photoionization is not energetically accessible (only bound photoionization, generating the parent cation, is possible). Thus any measured photofragment ions must originate from the decay of the resonances, and in this sense, studying the dissociaton of the ion-pair state probes the electron-ion resonances and their decay in the continuum in a clearer manner.

Coupling between ion-pair states and optically allowed continuum resonances has been studied in various molecular systems following photoabsorption and electron impact \cite{Dehmer,Dehmer1,Mitsuke,Kieffer,Oertel,Locht,Hao,Baklanov,Lucchese2,Lucchese1,Chang,Krauss,YTLee}. Optically forbidden continuum resonances, being more challenging to access, have remained largely unexamined both experimentally and theoretically. While they are not directly accessible by absorption of a single photon, optically dark resonances can be accessed directly by electron or ion impact excitation, and they are therefore relevant to the chemistry of planetary atmospheres and in technologies involving plasmas.

Such dark continuum states can be populated via two-photon absorption, which generally requires intense, ultrashort VUV pulses. If the bandwidth of the VUV excitation pulse is broad enough to span more than one member of a resonance progression, cation-anion pairs can emerge as dissociation products from different resonances. If the symmetries of two adjacent continuum resonances are different, the excitation pathways leading to them will involve differing combinations of parallel and perpendicular transitions. The symmetry of the resonances and their mode of excitation can be determined through energy- and angle-resolved photofragment ion distributions. As the relative amount of parallel and perpendicular transitions vary with the excitation energy, the photoion angular distribution relative to the polarization of the ionizing field will correspondingly vary with the kinetic energy release (KER), encoding the composition of orientations contributing to the two-photon transition and thus the symmetries of the resonances.

In this work, we present energy- and angle-resolved measurements of ion-pair production in O$_2$ following resonant two-photon absorption, where the O$^+$ ion is detected using 3-D momentum imaging. We use intense femtosecond 9.3 eV VUV pulses to populate ion-pair states of $^3\Sigma^-_g$ and $^3\Pi_g$ symmetry through excitation to optically dark, predissociated, continuum molecular Rydberg states. The variation in the KER-dependent photoion angular distribution maps the molecular orientations contributing to the two-photon transition, and indicates the presence of two narrowly separated continuum resonances of different symmetry that both couple to ion-pair states of the same total symmetry via internal conversion. 

\section{\label{sec:level2}Experiment}

The photodissociation dynamics of ion-pair formation in neutral O$_2$ molecules were investigated using the cold target recoil ion momentum imaging (COLTRIMS) technique \cite{Sturm}, where the cationic fragment produced by resonant two-photon absorption is collected with full 4$\pi$ solid angle and its 3-D momentum measured on an event-by-event basis. The positive ions are guided by parallel DC electric and magnetic fields (12.26 V/cm, 4.55 G) towards a position- and time-sensitive detector, consisting of a 120 mm multi-channel plate (MCP) chevron stack with a two-layer quad delay-line anode readout. Here, a charge carrier's 3-D momentum is encoded into its hit position on the detector and its time-of-flight relative to the laser trigger. The time-of-flight distinguishes O$^+$ fragment ions from O$_{2}^+$ parent cations by their mass-to-charge ratio. 

The laser system has been described in detail previously \cite{Sturm}, so we provide only a brief overview here, with a focus on recent changes made for the present experiments. A Ti:sapphire near-infrared (NIR) laser system produces 12~mJ, 45~fs pulses at a repetition rate of 50 Hz. The 800~nm pulses are frequency doubled using a 0.25 mm thick beta-barium borate crystal, where the copropagating NIR and blue fields are then separated using two dichroic mirrors. The reflected 400~nm photons ($\sim$3.6 mJ, $\sim$50 fs) are used to generate femtosecond VUV pulses via high harmonic generation (HHG) of the fundamental laser light, using a 6 m focusing geometry and a 10 cm long gas cell containing 3 Torr of krypton. The resulting VUV frequency comb is then separated from the 400~nm driving field by reflection from three Si mirrors near Brewster's angle for the fundamental, resulting in a suppression of the driving field by a factor of $<10^{-6}$. The 3rd harmonic (133 nm, 9.3 eV) is selected via transmission through a 0.30 mm thick MgF$_2$ VUV bandpass filter (Acton Optics FB130-B-1D.3), which totally suppresses the 5th harmonic and above. The femtosecond pulse duration of the 3rd harmonic is also maintained, while temporally separating the residual fundamental pulse from the 3rd harmonic pulse by roughly 850 fs, due to the difference in the group velocity dispersion (GVD) of the window at $\omega_0$ and $3\omega_0$ \cite{Allison,Li}. After transmission through the window, we estimate the pulse duration of the 3rd harmonic to be $\sim$35-40 fs, based on its spectral bandwidth, its estimated attochirp, and the thickness and GVD of the MgF$_2$ window \cite{Allison,Sekikawa1,Sekikawa2,Li}. The femtosecond 9.3 eV pulses are then back-focused (f = 15 cm) into the 3-D momentum imaging spectrometer using a protected Al mirror, the reflectance of which has been measured to be 43\% at 9.3eV \cite{Larsen}. The pulse energy of the 3rd harmonic on target is approximately 10 nJ, which was measured using a pair of bandpass VUV filters (same as above) and a calibrated photodiode. 

A vibrationally cold beam of oxygen molecules is prepared via an adiabatic expansion through a 0.03 mm nozzle, which is then collimated by a pair of skimmers. This molecular supersonic jet propagates perpendicular to the focusing VUV beam, where the two intersect in the interaction region of the spectrometer ($\sim$0.01 $\times$ 0.01 $\times$ 0.20 mm), resulting in a measured ionization rate of approximately 20~Hz by two-photon absorption in the target oxygen molecules. No significant ionization processes due to gas impurities, higher photon energies, or 3-photon transitions were detected. 

\section{\label{sec:level3}Theory}

Potential energy curves for the states of interest were computed using the MOLPRO program \cite{Molpro2012}. The one-electron basis set was aug-cc-pVTZ \cite{Dunning1989ccbasis,Kendall1992DunningAug}. Additional diffuse functions located at the bond midpoint were added to allow for an accurate representation of Rydberg orbitals \cite{Kaufmann1989RydBasis} including diffuse $s$-type Gaussians with exponents 0.024624, 0.01125,0.005858, 0.003346, and 0.0020484, $p$-type Gaussians with exponents 0.01925, 0.009988, 0.005689, 0.003476, and 0.002242, and diffuse $d$-type Gaussians with exponents of 0.06054, 0.02745, 0.01420, 0.008077, 0.004927. A full valence state averaged complete active space self-consistent field (SA-CASSCF) was computed including the $^3\Sigma_g^-$ ground state of the O$_2$ molecule and two $^3\Sigma_u^-$, two $^3\Delta_u$, and two of $^3\Pi_u$ symmetry. The orbital space included three $\sigma_g$,  two $\pi_u$, four $\sigma_u$, and one $\pi_g$ orbital. Using these orbitals, internally contracted multireference configuration interaction (ic-MRCI) calculations were performed including single and double excitations from the SA-CASSCF wave function.

\begin{figure}[h!]
    {
    \subfigure[]{%
        \includegraphics[width=8.0cm, trim=0.4cm 0.6cm 1.0cm 0.6cm, clip]{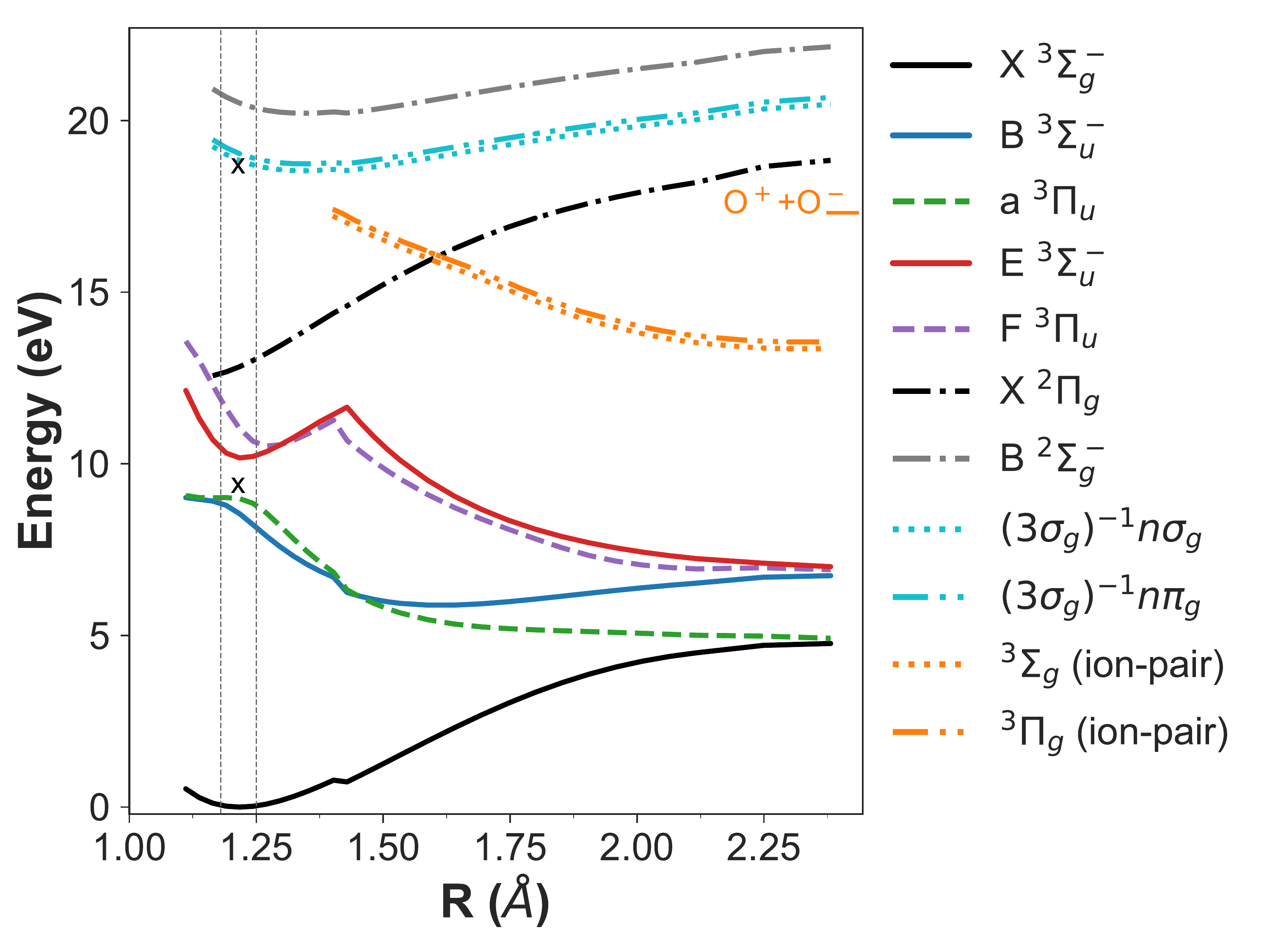}}}
    {
    \subfigure[]{%
        \includegraphics[width=8.0cm]{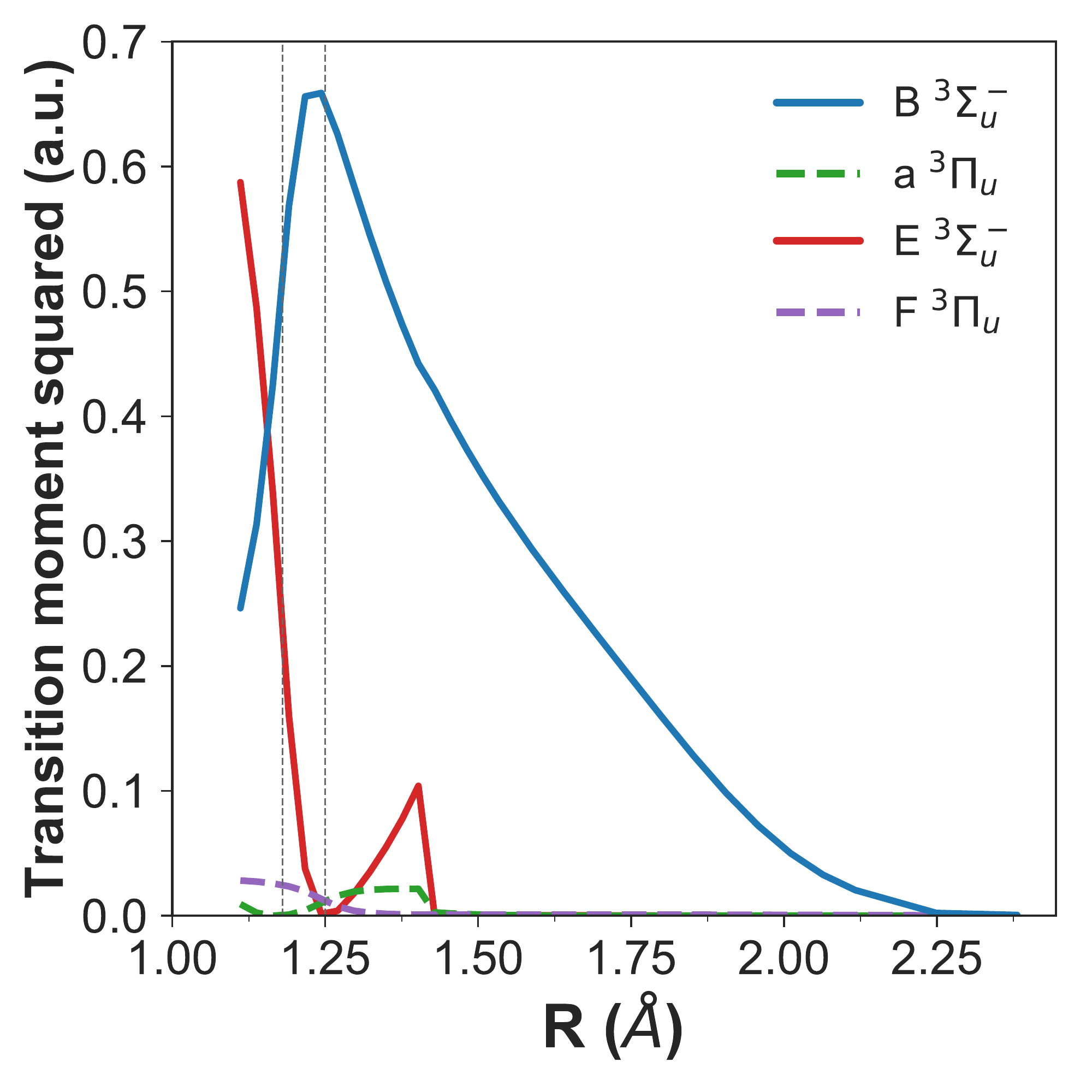}}}
\caption{(a) Potential energy curves for relevant states of molecular oxygen, with the one- and two-photon excitation energies indicated by the black x's. (b) The transition dipole moment squared for each of the excited states in the first excitation step. The FC region is indicated by the vertical dashed gray lines in (a) and (b). The dissociative limit for ion-pair production is indicated by the horizontal orange line at 17.3 eV on the right side of the figure in (a).}
\label{fig:PEC_tdm}
\end{figure}

The calculated potential energy curves (PECs) for the electronic states relevant to the following discussion are shown in Fig \ref{fig:PEC_tdm} (a). The ion-pair PECs are adapted from Ref.~\cite{Chang}. The ion-pair states in the Franck-Condon region, where crossings with the Rydberg states occur, are located at energies above the ground state of the ion. Thus they are meta-stable and can only be computed using a stabilization type calculation, which can be quite challenging at such a high energy. This type of calculation is beyond the scope of this paper. The X $^3\Sigma_g^-$ ground state of O$_2$ has the valence electron configuration (3$\sigma_g$)$^2$(1$\pi_u$)$^4$(1$\pi_g$)$^2$, while the two states with the dominant transition amplitudes (shown in Fig \ref{fig:PEC_tdm} (b)) are the  B $^3\Sigma_u^-$ and E $^3\Sigma_u^-$ states. These states have dominant electron configurations (3$\sigma_g$)$^2$(1$\pi_u$)$^3$(1$\pi_g$)$^3$ and (3$\sigma_g$)$^2$(1$\pi_u$)$^4$(1$\pi_g$)$^1$3p$\pi_u$, respectively, with configuration interaction between the two \cite{Buenker,Wang}. The electron configuration (3$\sigma_g$)$^1$(1$\pi_u$)$^4$(1$\pi_g$)$^2$(3$\sigma_u$)$^1$ also contributes to both states. 

Given the two-photon excitation energy of 18.6 eV ($\pm$ 0.15 eV), the relevant Rydberg series is that converging to the B $^2\Sigma_g^-$ ionic state, which has the valence electron configuration (3$\sigma_g$)$^1$(1$\pi_u$)$^4$(1$\pi_g$)$^2$n$\alpha$, where n is the principal quantum number of the Rydberg state and $\alpha$ is the molecular orbital, e.g. $\sigma_g$ and $\pi_g$. This Rydberg series is predissociated by ion-pair states of $^3\Sigma_g$ and $^3\Pi_g$ symmetry, which provide the continuum resonances a decay mechanism that competes with autoionization. These Rydberg states are of the opposite parity to those that have been previously studied in single-photon experiments and calculations, while the same is true for the ion-pair states that they couple to.

The intense VUV pulses in the present experiment are resonant with three electron configurations in O$_2$, therefore we consider two-photon excitation to occur in the following possible sequences. The first photon transition from the X $^3\Sigma_g^-$ to the B $^3\Sigma_u^-$ and E $^3\Sigma_u^-$ states involves a single electron excitation, either a $\pi_u \rightarrow \pi_g$, $\pi_g \rightarrow 3p\pi_u$, or $\sigma_g \rightarrow \sigma_u$ transition. Of the three electron configurations that result from these transitions, the first two do not provide a likely resonant pathway to the relevant Rydberg series, because a two-electron excitation is required to connect either of these intermediate state configurations to the relevant Rydberg series. Only the (3$\sigma_g$)$^1$(1$\pi_u$)$^4$(1$\pi_g$)$^2$(3$\sigma_u$)$^1$ configuration, resulting from the $\sigma_g \rightarrow \sigma_u$ transition, correlates with the electron configuration of the continuum resonances associated with the B $^2\Sigma_g^-$ state, in the sense that they connect through a single electron excitation. In this case, Rydberg states with the configurations (3$\sigma_g$)$^1$(1$\pi_u$)$^4$(1$\pi_g$)$^2n\sigma_g$ and (3$\sigma_g$)$^1$(1$\pi_u$)$^4$(1$\pi_g$)$^2n\pi_g$
can be populated via a parallel $\sigma_u \rightarrow \sigma_g$ or perpendicular $\sigma_u \rightarrow \pi_g$ transition, respectively. 

From this process of deduction, we arrive at two distinct two-photon pathways that connect with the target continuum Rydberg states, namely:

\begin{equation}
  (3\sigma_g)^2(1\pi_u)^4(1\pi_g)^2 \rightarrow (3\sigma_g)^{-1}3\sigma_u \rightarrow (3\sigma_g)^{-1}n\sigma_g  
\end{equation}
\begin{equation}
  (3\sigma_g)^2(1\pi_u)^4(1\pi_g)^2 \rightarrow (3\sigma_g)^{-1}3\sigma_u \rightarrow (3\sigma_g)^{-1}n\pi_g
\end{equation}

where pathway (1) is a parallel-parallel ($\parallel$-$\parallel$) transition, $\sigma_g \rightarrow \sigma_u \rightarrow \sigma_g$, and pathway (2) is a parallel-perpendicular ($\parallel$-$\perp$) transition, $\sigma_g \rightarrow \sigma_u \rightarrow \pi_g$. Since the intermediate state electron configuration contributes to both the B $^3\Sigma_u^-$ and E $^3\Sigma_u^-$ states and the excitation energy is in an avoided crossing region, the first transition creates a superposition of these two electronic states. The second photon then projects the excitation onto a continuum resonance of (3$\sigma_g$)$^{-1}$n$\sigma_g$ or (3$\sigma_g$)$^{-1}$n$\pi_g$ Rydberg character. Both Rydberg states can decay by autoionization, or through internal conversion to an ion-pair state with the same symmetry.

\begin{figure}[h!]
    {\includegraphics[width=8.0cm]{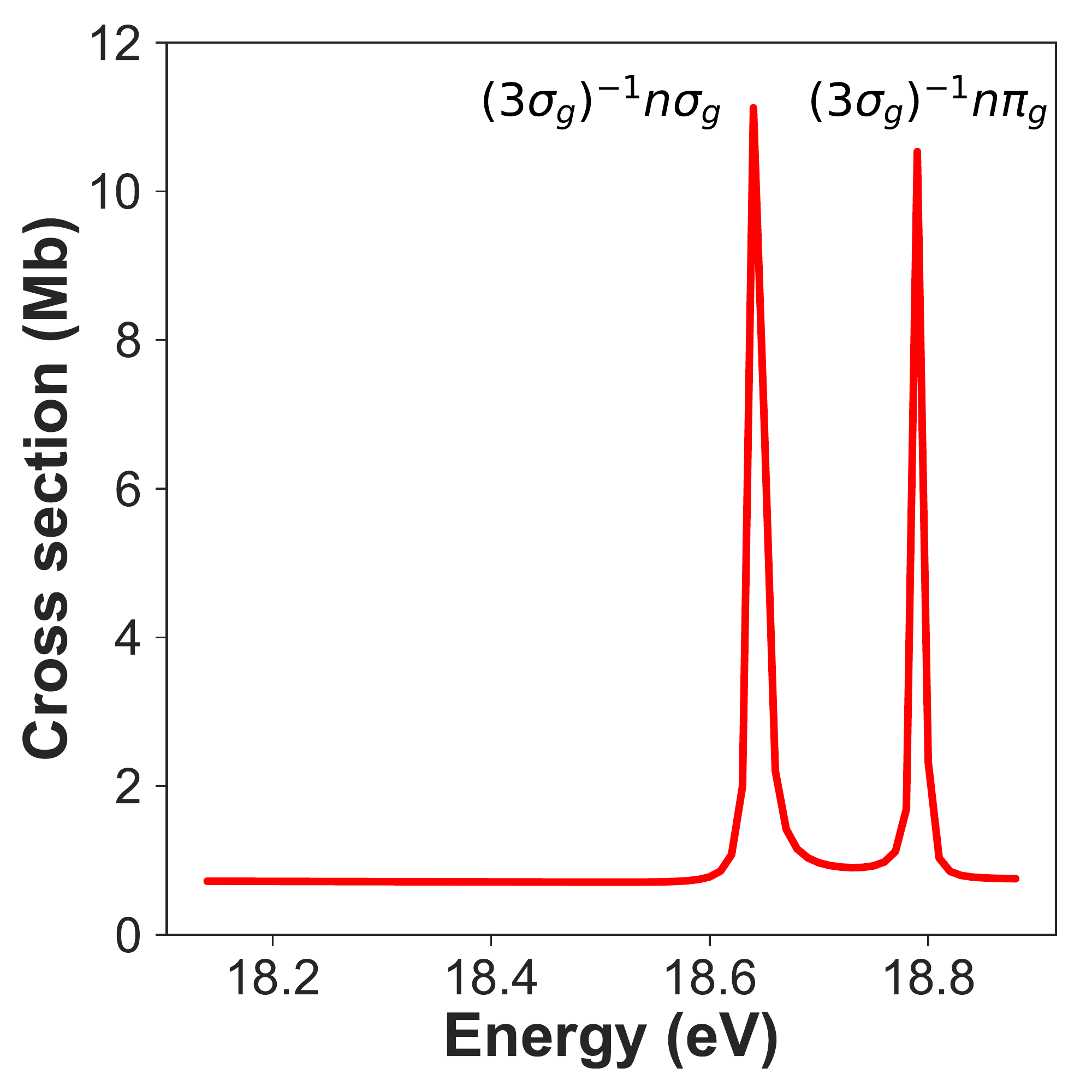}}
\caption{The calculated photoionization cross section in the region near the (3$\sigma_g$)$^{-1}n\sigma_g$ and (3$\sigma_g$)$^{-1}n\pi_g$ autoionizing resonances.}
\label{fig:resonances}
\end{figure}

One-photon photoionization calculations were performed at a fixed geometry with $R_{\textrm{O-O}}=1.19\,\textrm{\AA}$ using the Schwinger variational method \cite{Stratmann1995,Stratmann1996}. The initial state for the excitation to the autoionizing Rydberg state was the B $^3\Sigma_u^-$ state, which at this geometry also has a significant Rydberg character. The autoionizing state was written as a close coupling expansion of antisymmeterized products of continuum orbitals times a set of ion states, which included the X $^2\Pi_g$, a $^4\Pi_u$, A $^2\Pi_u$, b~$^4\Sigma_g^-$ and B~$^2\Sigma_g^-$ ion states. The ion states and the initial state were computed using the orbitals from the SA-CASSCF calculations disussed above. In each state a full valence configuration interaction calculation was performed using the orbitals from the SA-CASSCF calculations. The positions of these two autoionizing states in Fig.~\ref{fig:PEC_tdm}~(a), shown as the cyan PECs, stem from combining the quantum defect computed in the Schwinger calculation with the ion state energy of the MRCI calculation.

The cross sections of the two calculated resonances are shown in Fig~\ref{fig:resonances}. The Fano line shapes of these two autoionizing resonances are fit in accordance with the equation Fano proposed in 1961 for a resonance interacting with two or more continua \cite{Fano},

\begin{equation}
  \sigma(\omega) = A + B \left[\frac{(q+\epsilon(\omega))^2}{1+\epsilon(\omega)^2}\right]
\label{eq:FanoLS}
\end{equation}
where $\epsilon(\omega) = (\omega - \omega_0)/(\Gamma/2)$. The $q$ parameter, width $\Gamma$, and position $\omega_0$ for the Fano profile fits of the two resonances are displayed in Table \ref{table:Fano}. Note the different sign of $q$ for each resonance, which is consistent with the fact that each resonance is a member of a different Rydberg series that belong to continua of different symmetries. Since $q^2 \gg 1$ for both resonances, we conclude that direct two-photon coupling is the primary mechanism for populating the two continuum resonances, rather than excitation to the continuum combined with coupling between the continuum and discrete resonances ("intensity borrowing") \cite{Plunkett}. We estimate the lifetime of each resonance from its energy width $\Gamma$, using the time-energy uncertainty principle, finding lifetimes of 8~fs and 11~fs for the (3$\sigma_g$)$^{-1}$n$\sigma_g$ and (3$\sigma_g$)$^{-1}$n$\pi_g$ resonances, respectively. 

\begin{table}[h!]
\centering
\begin{tabular}{  c  c  c  c  } 
 \hline
 Resonance & $q$ & $\Gamma$ (meV) & $\omega_0$ (eV)\\
 \hline
 (3$\sigma_g$)$^{-1}$n$\sigma_g$ & 28.76 & 82 & 18.64 \\
 (3$\sigma_g$)$^{-1}$n$\pi_g$ & -45.02 & 58 & 18.79 \\
 \hline
\end{tabular}
\caption{The Fano line shape fit parameters $q$, $\Gamma$, and $\omega_0$ for the two calculated resonances.}
\label{table:Fano}
\end{table}

\section{\label{sec:level4}Results}

The calculated 130 - 135 nm VUV photoabsorption partial cross sections of Ref.~\cite{Dalgarno} for excitation to the B~$^3\Sigma^-_u$ and E~$^3\Sigma^-_u$ states are more than an order of magnitude larger than those of the a~$^3\Pi_u$ and F~$^3\Pi_u$ states at the equilibrium internuclear distance, which is consistent with our calculations shown in Fig~\ref{fig:PEC_tdm}~(b). Thus at these photon energies parallel dipole transitions are strongly favored over perpendicular transitions. Using a sufficiently intense VUV pulse, the B~$^3\Sigma^-_u$ and E~$^3\Sigma^-_u$ states can be resonantly populated, where a fraction of the excited population can then undergo a second transition driven within the same VUV pulse. In this way, two 9.3 eV photons can access parity-forbidden molecular Rydberg states converging to the B~$^2\Sigma^-_g$ state of O$_2^+$. This second photon transition may be either parallel or perpendicular. These discrete continuum states are predissociated by ion-pair states \cite{Baklanov,Oertel,Dehmer,Hao,Chang,Krauss,Mitsuke}, enabling the molecular orientation during photoabsorption to be captured and by extension the symmetries of the continuum resonances participating. Since this two-photon energy lies below the dissociative ionization threshold \cite{Blyth}, any oxygen cations produced must originate from the decay of these continuum resonances into ion-pair channels.

\begin{figure}[h!]
    {
    \subfigure[]{%
        \includegraphics[width=8.0cm]{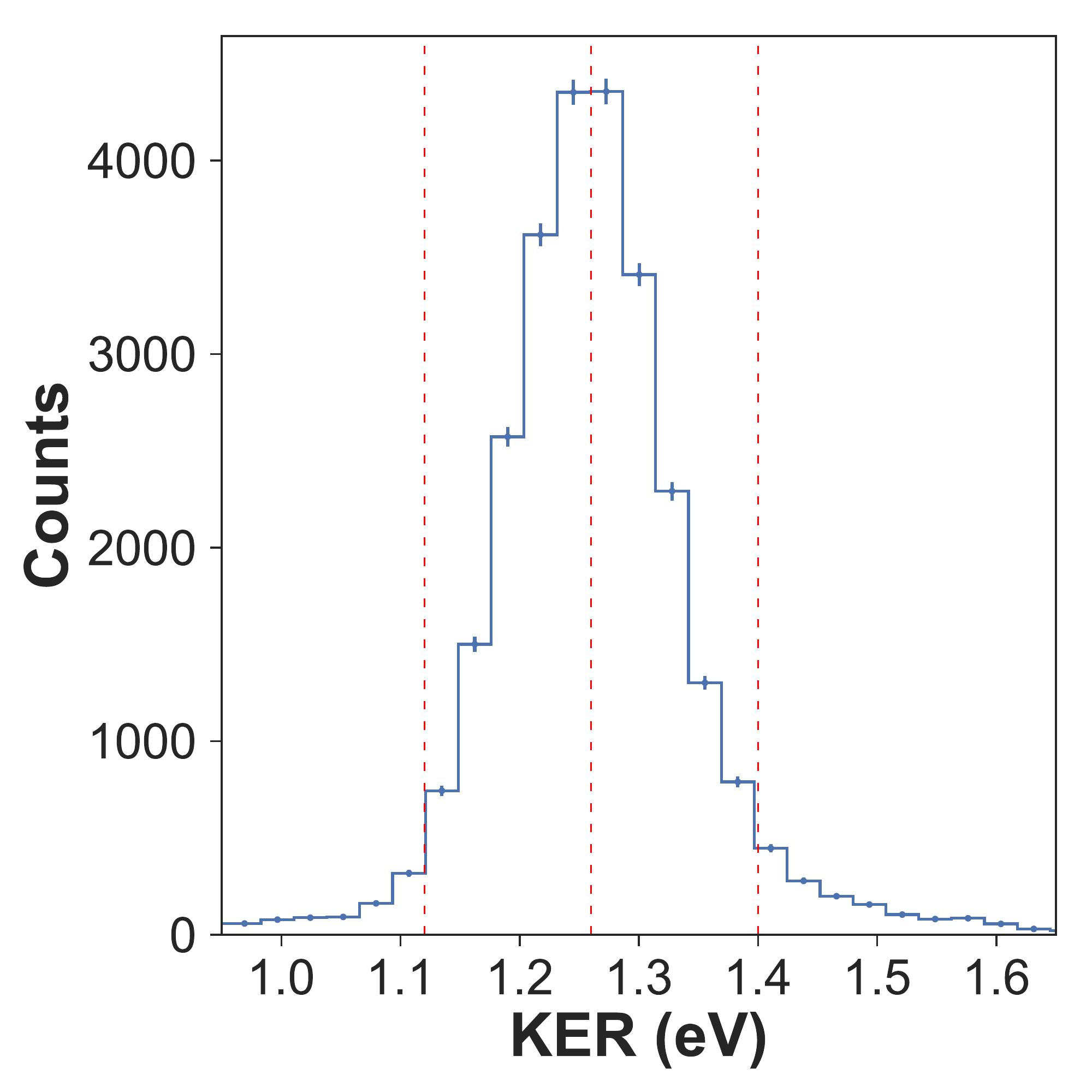}}}
    {
    \subfigure[]{%
        \includegraphics[width=8.0cm]{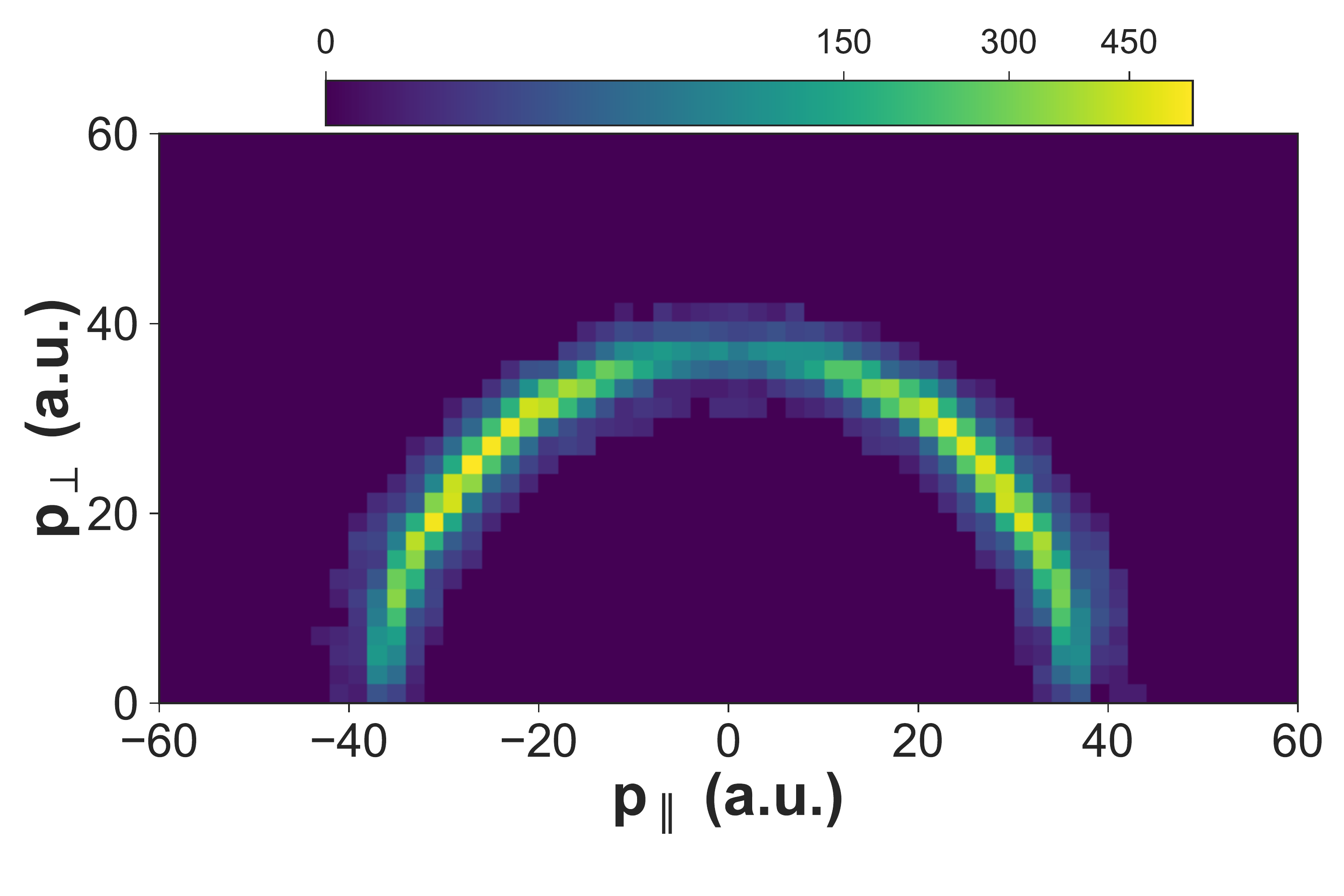}}}
\caption{(a) The ion-pair production KER and (b) O$^+$ momentum distribution parallel versus perpendicular to the VUV polarization. The red dashed vertical lines in (a) indicate two different regions of the KER distribution that are subsequently analyzed.}
\label{fig:KER_pParaTrans}
\end{figure}

The branching fractions for dissociative ion-pair production (O$^+$-O$^-$) versus bound two-photon ionization (O$_2^+$) is measured to be roughly 22\% and 78\%, respectively. The adiabatic dissociation limit of the ion-pair states is $\sim$17.3 eV~\cite{Krauss}, thus for an excitation energy of 18.6~eV we expect approximately 1.3~eV of KER. The measured KER of the dissociating ion-pair is presented in Fig.~\ref{fig:KER_pParaTrans}~(a). We observe a single peak centered near 1.26 eV, with a full width at half maximum (FWHM) of $\sim$150~meV, which is considerably smaller than the two-photon bandwidth ($\sim$300~meV) of the excitation pulse. We attribute the narrow KER to a resonant two-photon excitation of a single vibrational level of either of two predissociated Rydberg states that are nearly degenerate in energy. The width of the KER distribution corresponds with the widths and spacing of these two narrowly separated resonances, convoluted with the energy resolution of the ion detector, rather than the two-photon bandwidth of the excitation, as the ion-pair states are not directly accessed via a 2-photon transition. According to the reflection approximation, if the dissociative ion-pair states were excited directly, the width of the KER would correspond with the two-photon bandwidth, whereas the measured KER distribution is significantly narrower. Although two resonance peaks can not be resolved in the measured KER alone, we find that the two neighboring continuum states can be distinguished by their symmetry, where different regions of the KER distribution exhibit varying photoion angular distributions. The O$^+$ photoion momentum distribution transverse versus parallel to the VUV polarization vector is shown in Fig~\ref{fig:KER_pParaTrans}~(b). Here p$_\perp$ and p$_\parallel$ are the ion momentum components transverse and parallel to the VUV polarization, respectively. The photoion momentum distribution tends to exhibit high longitudinal momentum and low transverse momentum. To differentiate the two continuum resonances by their symmetries, we turn to the photoion angular distribution \cite{zareDopplerLineShape1963,dunnAnisotropiesAngularDistributions1962}.

\begin{figure}[h!]
\includegraphics[width=8.0cm]{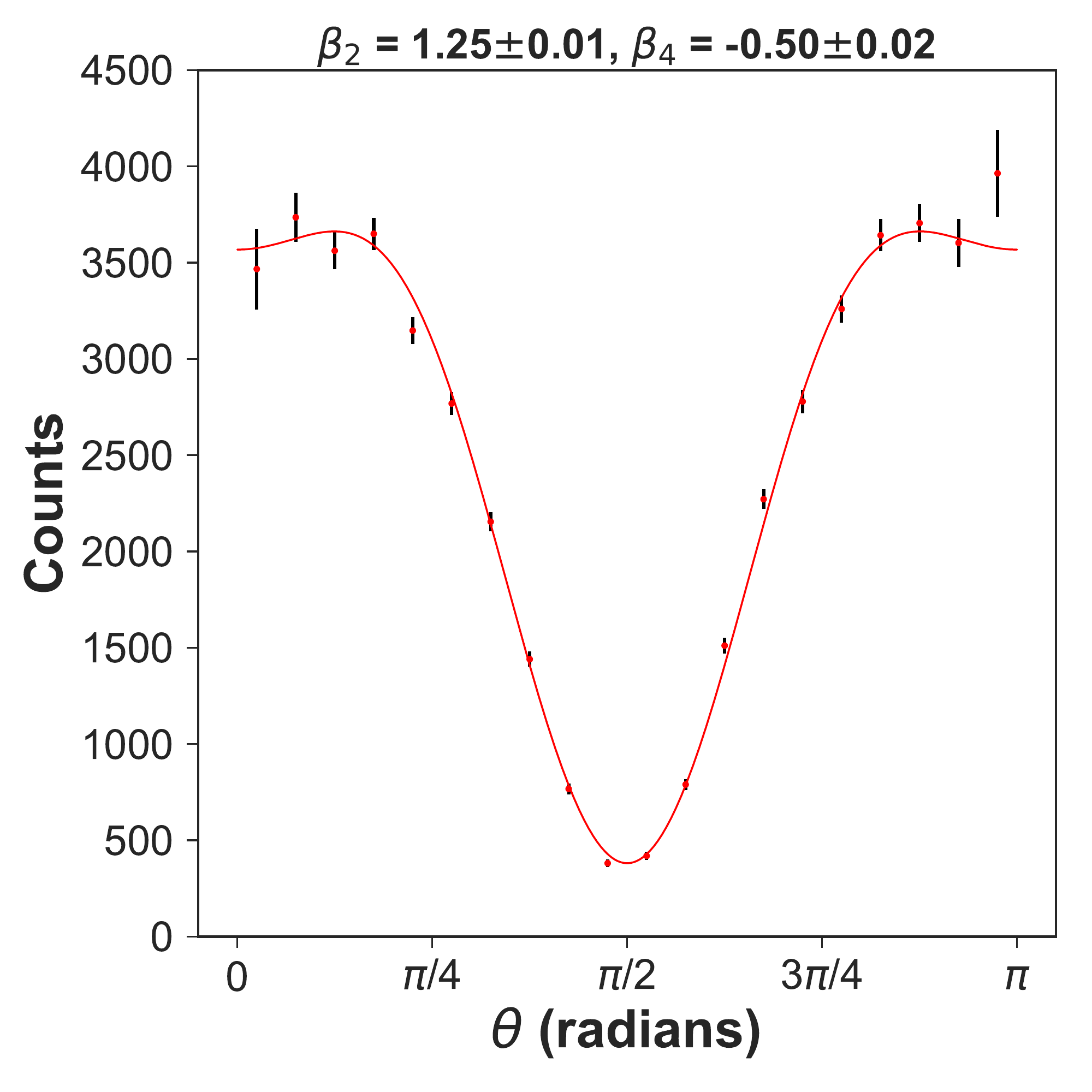}
\caption{The energy-integrated O$^+$ photoion angular distribution from ion-pair formation in O$_2$ following resonant two-photon absorption at 9.3 eV. The data is fit using equation (1), where the retrieved $\beta$ parameters are displayed above the plot.}
\label{fig:KERTheta}
\end{figure}

For two-photon processes, the energy-dependent photofragment angular distribution of a molecule dissociated by linearly polarized light is given by the parameterization:

\begin{equation}
    \frac{d^2\sigma}{d\Omega dE} = \frac{\sigma_0(E)}{4\pi} [1 + \beta_{2}(E)P_{2}(\cos\theta) + \beta_{4}(E)P_{4}(\cos\theta)]
\label{eq:DDCS}
\end{equation}

where $\sigma_0$ is the total cross section, $\theta$ is the angle between the photoion momentum vector and the polarization vector of the light, E is the kinetic energy release, $\beta_2$ and $\beta_4$ are the second and fourth order anisotropy parameters, and $P_2$ and $P_4$ are the second and fourth order Legendre polynomials in variable $\cos\theta$. The measured energy-integrated angle-differential photoion distribution is presented in Fig \ref{fig:KERTheta}, where the data has been fit (solid red line) in accordance with equation \ref{eq:DDCS}, using the projection method discussed in \cite{Lucchese0}, and the error on the $\beta$ parameters determined via statistical bootstrapping \cite{Efron}. The $\beta$ parameters retrieved from the fit are $\beta_2$ = 1.25$\pm$0.01, $\beta_4$ = -0.50$\pm$0.02. 

\begin{figure}[h!]
    {
    \subfigure[]{%
        \includegraphics[width=8.0cm]{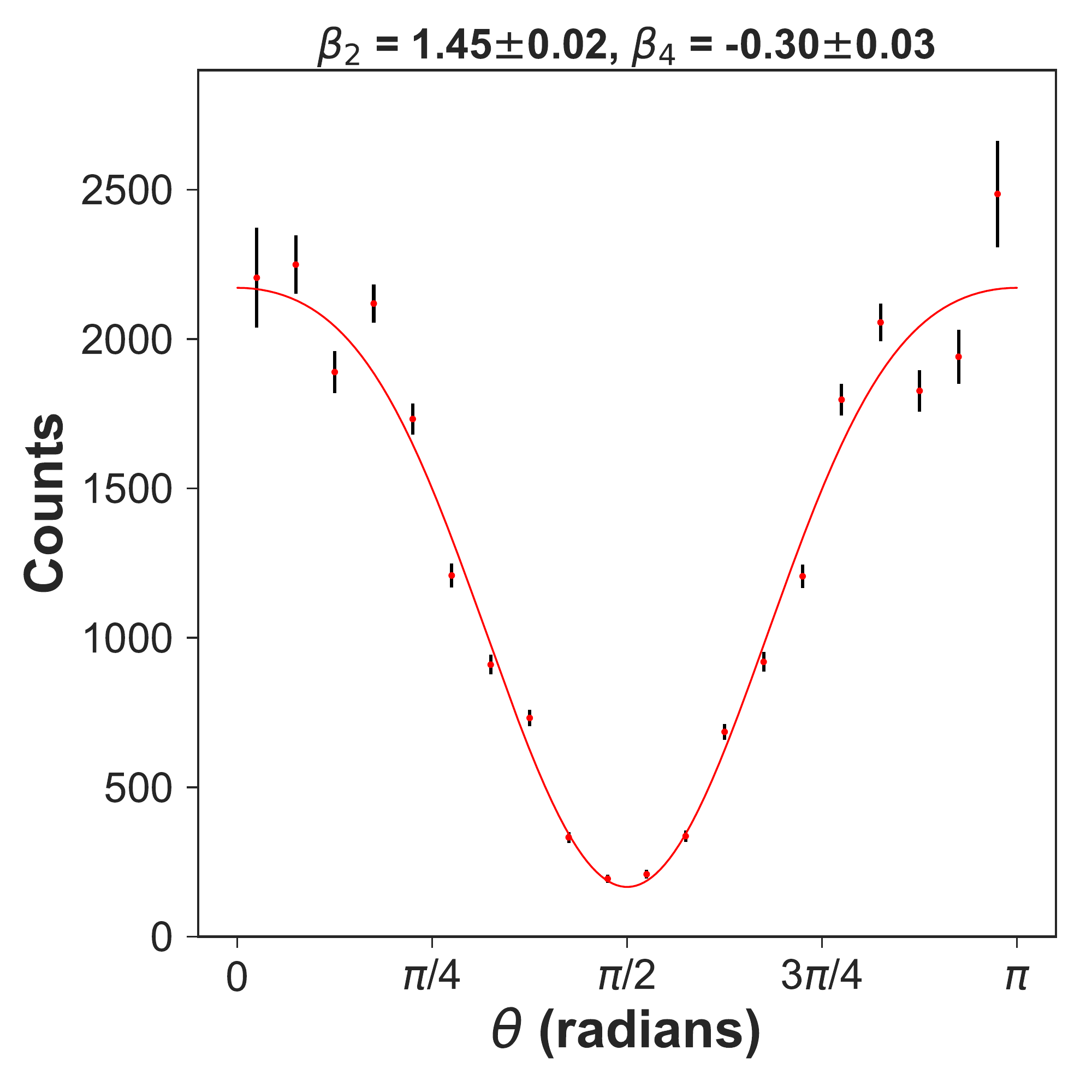}}}
    {
    \subfigure[]{%
        \includegraphics[width=8.0cm]{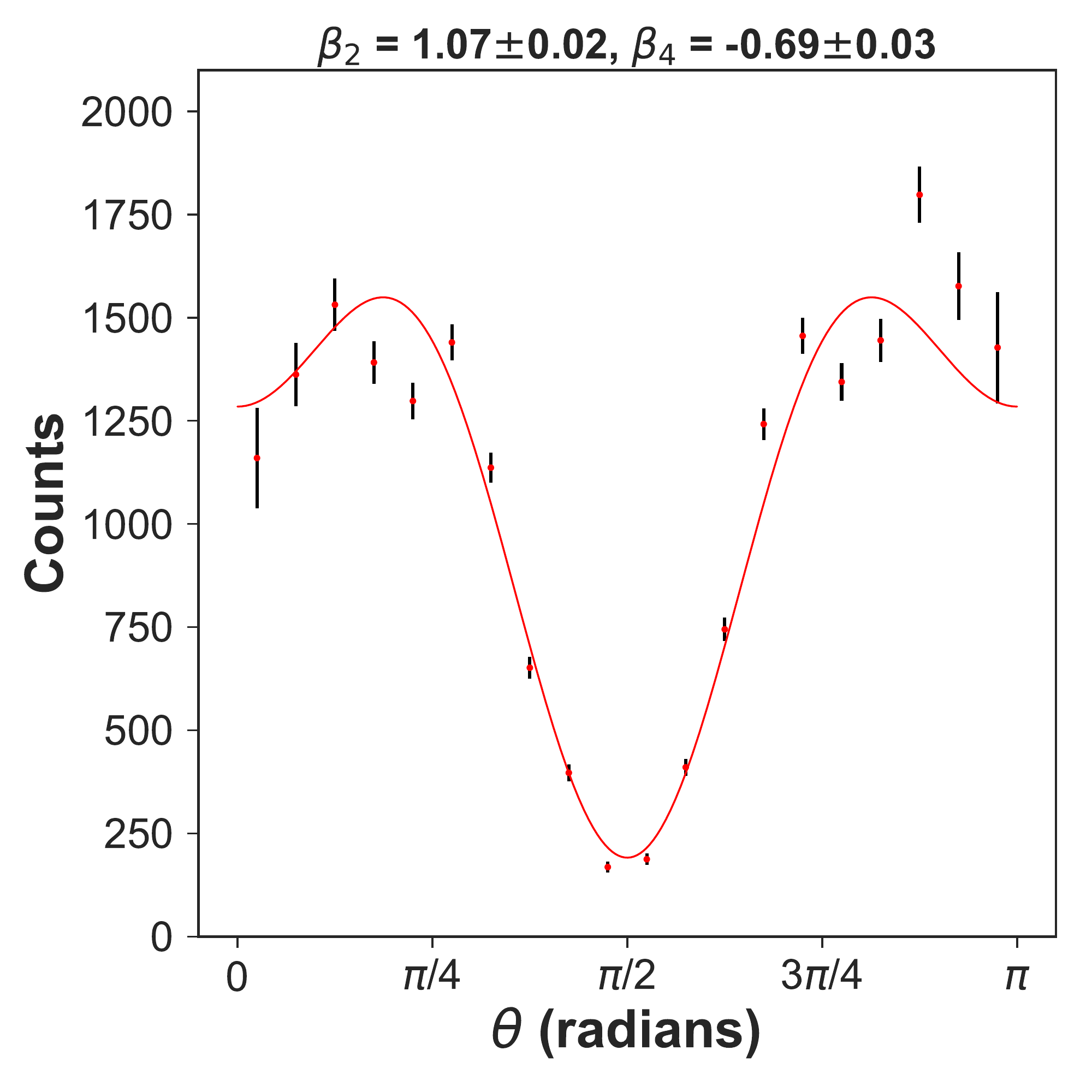}}}
\caption{The energy-resolved photoion angular distribution for the O$^+$ fragment produced from ion-pair formation in O$_2$ following resonant two-photon absorption at 9.3 eV for the low KER region in (a) and the high KER region in (b). These two KER regions are indicated in Fig \ref{fig:KER_pParaTrans} (a). The data is fit in accordance with equation (1), where the retrieved $\beta$ parameters are shown above each plot.}
\label{fig:HighLowKERTheta}
\end{figure}

The negative value of $\beta_4$ indicates that there is a $\parallel$-$\perp$ two-photon transition that connects to the ion-pair formation states, since $\beta_4$ is positive for $\parallel$-$\parallel$ and $\perp$-$\perp$ transitions \cite{Dixon}. The varying contribution from the $\parallel$-$\perp$ pathway can be probed by considering different regions of the KER distribution. This is because different regions of the KER distribution correspond with different two-photon excitation energies. As the excitation energy is varied, the scattering cross section for different symmetry autoionizing states can come in and out of resonance. This will lead to a corresponding variation in the $\beta$ parameters. We divide the momentum distribution into two KER regions, indicated by the red vertical dashed lines in Fig \ref{fig:KER_pParaTrans} (a). These two slices are taken over the KER intervals 1.12 - 1.26 eV and 1.26 - 1.40 eV, where the centers of the two regions are separated by 0.14 eV in energy. The corresponding O$^+$ photofragment angular distributions for each KER slice and their $\beta$ parameters are shown in Fig \ref{fig:HighLowKERTheta}. 

The energy-resolved $\beta$ parameters vary for the two KER slices, with $\beta_2$ = 1.45$\pm$0.02 and $\beta_4$ = -0.30$\pm$0.03 for the low KER region, and $\beta_2$ = 1.07$\pm$0.02 and $\beta_4$ = -0.69$\pm$0.03 for the high KER region. Here, as the KER increases, both $\beta$ values decrease, meaning an increase in the relative amount of $\parallel$-$\perp$ two-photon transitions. Since the high energy region of the KER distribution corresponds with higher excitation energies, the photoion angular distribution indicates the existence of two narrowly spaced continuum resonances possessing different symmetries, one of dominant (3$\sigma_g$)$^{-1}n\sigma_g$ character and the other with dominant (3$\sigma_g$)$^{-1}$n$\pi_g$ character lying slightly higher in energy. As the energy of the excitation increases, the scattering amplitude of the higher lying resonance increases while that of the lower lying resonance decreases and thus the amount of population decaying through this (3$\sigma_g$)$^{-1}n\pi_g$ resonance via the $\parallel$-$\perp$ pathway increases and the population decaying through the (3$\sigma_g$)$^{-1}n\sigma_g$ resonance via the $\parallel$-$\parallel$ pathway decreases. This is reflected by the energy-resolved photoion angular distributions, while the width of the KER distribution indicates these two resonances are nearly degenerate in energy, with a spacing of $\sim$100 meV. We point out that this analysis is predicated on the validity of the axial recoil approximation, i.e. prompt dissociation before significant rotation of the molecule. Since the observed dissociation requires a non-adiabatic transition from an autoionizing state to an ion-pair state, this can potentially lead to a violation of this approximation, if the resonance lifetime is not short-lived and comparable to the rotational period of the molecule. However, we note that the lifetimes of the autoioning states are estimated to be much less than 100~fs. Furthermore, given the considerable anisotropy in the measured angular distributions, and significant changes in the anisotropy parameters across the narrow KER distribution, the non-adiabatic coupling apparently occurs on a timescale that is considerably faster than the rotational motion of the molecule, confirming the axial recoil approximation as valid. 

\section{\label{sec:level5}Conclusion}

In the experiments reported here, we used intense femtosecond 9.3~eV VUV pulses, resonant within the Schumann-Runge continuum, and 3-D momentum imaging to study the decay of metastable, optically dark, continuum resonances in molecular oxygen, which are populated following two-photon absorption. These resonances are parity-forbidden in single-photon experiments, and to our knowledge, have not been studied before experimentally or theoretically. Since scattering through such resonances often weakly competes with the stronger direct ionization process, studying them in detail is challenging. Here, we overcame some of these challenges through the use of photofragment ion detection. A dipole allowed parallel transition to the B~$^3\Sigma^-_u$ state is excited by a femtosecond VUV pulse. This valence states mixes with the E~$^3\Sigma^-_u$ Rydberg state. A second VUV photon from within the same pulse populates continuum resonances lying below the dissociative ionization threshold, which can decay through to the ion-pair states, interrupting and competing with autoionization. The KER-dependent photoion angular distribution emerges from two optically dark, narrowly separated continuum resonances of different symmetry and reveals that these resonances are excited by $\parallel$-$\parallel$ and $\parallel$-$\perp$ two-photon transitions, where both decay by coupling to the ion-pair formation states of the same total symmetry via internal conversion. This work demonstrates that energy- and angle-resolved fragment ion momentum imaging can be highly sensitive to the symmetry of predissociating electronic states in the ionization continuum.

\section{\label{sec:level6}Acknowledgments}

This research was supported by the DOE, Office of Science, Office of Basic Energy Sciences, Chemical Sciences, Geosciences, and Biosciences Division, under Contract No. DE-AC02-05CH11231.  This research used the resources of the National Energy Research Scientific Computing Center, a DOE Office of Science User Facility, and the Lawrencium computational cluster resource provided by the IT Division at the Lawrence Berkeley National Laboratory.

\section{\label{sec:level7}Data Availability}

The data supporting the findings of this study are available from the corresponding authors upon request.

\bibliography{O2_ionpair_refs}

\end{document}